\documentclass{appolb}
\usepackage{epsfig}
\begin{document}
\title{GLOBAL VARIABLES FOR VARIOUS CENTRALITIES AT RHIC: A CRACOW MODEL APPROACH
\thanks{Presented at the XLVI Cracow School of Theoretical Physics, Zakopane,
Poland,\\ May 27-June 5, 2006}%
}
\author{Dariusz PROROK
\address{Institute of
Theoretical Physics, University of Wroc{\l}aw,\\ Pl.Maksa Borna 9,
50-204 Wroc{\l}aw, Poland} } \maketitle
\begin{abstract}
The final $p_{T}$-spectra measured at RHIC at $\sqrt{s_{NN}}=130$
and $200$ GeV are fitted within the Cracow single-freeze-out
model. Then the global variables like the transverse energy at
midrapidity, the charged particle multiplicity at midrapidity and
the total multiplicity of charged particles are evaluated. The
predictions agree fairly well with the experimental data. The
centrality independence of the total number of charged particles
per participant pair has been also reproduced.
\end{abstract}

\section{The single-freeze-out model and global variable estimates}
In this lecture the application of the single-freeze-out
statistical model
\cite{Florkowski:2001fp,Broniowski:2001we,Broniowski:2001uk,Broniowski:2002nf,Baran:2003nm}
to the final data on the $p_{T}$ spectra of identified charged
particles measured in the Relativistic Heavy Ion Collider (RHIC)
at $\sqrt{s_{NN}}=130$ and 200 GeV
\cite{Adcox:2001mf,Adcox:2003nr,Adler:2003cb,Adams:2003xp,Arsene:2005mr,Bearden:2004yx,Bearden:2003hx}
is reviewed. Details of this analysis can be found elsewhere
\cite{Prorok:2005uv}. The foundations of the model are as follows:
(\textit{a}) the chemical and thermal freeze-outs take place
simultaneously, (\textit{b}) all confirmed resonances up to a mass
of $2$ GeV from the Particle Data Tables \cite{Hagiwara:fs} are
taken into account, (\textit{c}) a freeze-out hypersurface is
defined by the equation $\tau =
\sqrt{t^{2}-r_{x}^{2}-r_{y}^{2}-r_{z}^{2}}= const$, (\textit{d})
the four-velocity of an element of the freeze-out hypersurface is
proportional to its coordinate, $u^{\mu}= x^{\mu} / \tau$,
(\textit{e}) the transverse size is restricted by the condition
$r=\sqrt{r_{x}^{2}+r_{y}^{2}}< \rho_{max}$. The maximum
transverse-flow parameter can be expressed as
$\beta_{\perp}^{max}=(\rho_{max}/\tau)
/(\sqrt{1+(\rho_{max}/\tau)^{2}})$. The model has four parameters,
namely, the two thermal parameters, the temperature $T$ and the
baryon number chemical potential $\mu_{B}$, and the two geometric
parameters, $\tau$ and $\rho_{max}$.

With the use of the following parametrization of the hypersurface

\begin{eqnarray}
t= \tau \cosh{\alpha_{\parallel}}\cosh{\alpha_{\perp}},\;\;\;
r_{x}=  \tau \sinh{\alpha_{\perp}}\cos{\phi}, \cr r_{y}=  \tau
\sinh{\alpha_{\perp}}\sin{\phi},\;\;\;r_{z}=\tau
\sinh{\alpha_{\parallel}}\cosh{\alpha_{\perp}}, \label{Parahyp}
\end{eqnarray}

\noindent the invariant distribution of the measured particles of
species $i$ can be expressed in the form

\begin{equation}
{ {dN_{i}} \over {d^{2}p_{T}\;dy} }= \tau^{3}\;
\int\limits_{-\infty}^{+\infty}
d\alpha_{\parallel}\;\int\limits_{0}^{\rho_{max}/\tau}\;\sinh{\alpha_{\perp}}
d(\sinh{\alpha_{\perp}})\; \int\limits_{0}^{2\pi} d\xi\;(p \cdot
u) \; f_{i}(p \cdot u) \;, \label{Cooper2}
\end{equation}

\noindent where

\begin{equation}
p \cdot u =
m_{T}\cosh{(\alpha_{\parallel}-y)}\cosh{\alpha_{\perp}}-
p_{T}\cos{\xi}\sinh{\alpha_{\perp}}\;, \label{Peu}
\end{equation}

\noindent and $f_{i}$ is the final momentum distribution of the
particle in question. The final distribution means here that
$f_{i}$ is the sum of primordial and simple and sequential decay
contributions to the particle distribution (for details see
\cite{Broniowski:2002nf,Prorok:2004af}).

Having integrated the distribution, Eq.~(\ref{Cooper2}), over
$p_{T}$ and summing over appropriate final particles, one can
obtain transverse energy ($dE_{T}/d\eta\vert_{mid}$) and charged
particle multiplicity ($dN_{ch}/d\eta\vert_{mid}$) densities at
mid-rapidity (for details see \cite{Prorok:2005uv}). The
experimentally measured transverse energy is defined as

\begin{equation}
E_{T} = \sum_{i = 1}^{L} \hat{E}_{i} \cdot \sin{\theta_{i}} \;,
\label{Etdef}
\end{equation}

\noindent where $\theta_{i}$ is the polar angle, $\hat{E}_{i}$
denotes $E_{i}-m_{N}$ ($m_{N}$ means the nucleon mass) for
baryons, $E_{i}+m_{N}$ for antibaryons and the total energy
$E_{i}$ for all other particles, and the sum is taken over all $L$
emitted particles \cite{Adams:2004cb,Adler:2004zn}.

In the case of expansion satisfying the condition $d\sigma_{\mu}
\sim u_{\mu}$ on a freeze-out hypersurface (as here), the total
multiplicity of particle species $i$ can be derived in the form

\begin{eqnarray}
N_{i} &=& \int d^{2}p_{T}\;dy\;{{dN_{i}} \over {d^{2}p_{T}\;dy}}=
\int d\sigma \int d^{2}p_{T}\;dy\;(p \cdot u)\;f_{i}(p \cdot u)
\cr \cr && = \int d\sigma \; n_{i}(T,\mu_{B}) =
n_{i}(T,\mu_{B})\int d\sigma \;, \label{Totmult}
\end{eqnarray}

\noindent if the local thermal parameters are constant on this
hypersurface. The density $n_{i}$ is not the primordial thermal
density of particle species $i$ but it collects also contributions
from decays of resonances. The last integral in
Eq.~(\ref{Totmult}) expresses the hypersurface volume and if the
rapidity of the fluid element $\alpha_{\parallel}$ is unlimited
(\cf Eq.~(\ref{Cooper2})), then this volume will be infinite. Thus
$\alpha_{\parallel}$ is assumed to have its maximal value
$\alpha_{\parallel}^{max}$. Then the volume can be expressed as
$2\pi\;\alpha_{\parallel}^{max} \tau \rho_{max}^{2}$ and the total
multiplicity of charged particles is obtained:

\begin{equation}
N_{ch} = 2\pi\;\alpha_{\parallel}^{max} \tau \rho_{max}^{2}
\sum_{i \in B} n_{i}(T,\mu_{B}) = 2\pi\;\alpha_{\parallel}^{max}
\tau \rho_{max}^{2}\; n_{ch}(T,\mu_{B}) \;,\label{Totcharged}
\end{equation}

\noindent where $B=\{\pi^{+},\; \pi^{-},\; K^{+},\; K^{-},\; p,\;
\bar{p}\}$. For $\alpha_{\parallel}^{max}$ the following
parametrization is obtained (for details see
\cite{Prorok:2005uv}):

\begin{equation}
\alpha_{\parallel}^{max}(c) = y_{p} - { \langle \delta y \rangle
\over 0.975 } \cdot (1-c) \;,\label{Alfpmax}
\end{equation}

\noindent where $y_{p}$ is the projectile rapidity, $\langle
\delta y \rangle$ the average rapidity loss and $c$ is a
fractional number representing the middle of a given centrality
bin, \emph{i.e.} $c=0.025$ for the $0-5 \%$ centrality bin, etc..

\section{Results}

Studies of the particle ratios and $p_{T}$ spectra at various
centralities in the framework of the single freeze-out model were
done for the preliminary RHIC data at $\sqrt{s_{NN}}=200$ GeV
\cite{Chujo:2002bi,Barannikova:2002qw} in \cite{Baran:2003nm}. The
procedure has two stages. First, thermal parameters $T$ and
$\mu_{B}$ are fitted with the use of the experimental ratios of
hadron multiplicities at midrapidity. After then two next
parameters, $\tau$ and $\rho_{max}$, are determined from the
simultaneous fit to the transverse-momentum spectra of
$\pi^{\pm}$, $K^{\pm}$, $p$ and $\bar{p}$. Both fits are performed
with the help of the $\chi^{2}$ method. Since (\textit{a}) the
preliminary data for the $p_{T}$ spectra
\cite{Chujo:2002bi,Barannikova:2002qw} differ from the final data
\cite{Adler:2003cb,Adams:2003xp}, (\textit{b}) not all bins were
fitted in \cite{Baran:2003nm}, (\textit{c}) the data points were
digitized from the plots in \cite{Baran:2003nm}, the fit procedure
for determination of the geometric parameters of the model, $\tau$
and $\rho_{max}$, has been performed again. For completeness the
PHENIX case at $\sqrt{s_{NN}}=130$ GeV has been repeated, since
the first published data were for three bins only
\cite{Adcox:2001mf}. The data for the next two bins, which were
not fitted in \cite{Broniowski:2002nf}, were added in the later
report \cite{Adcox:2003nr}. Also the BRAHMS spectra measured at
$\sqrt{s_{NN}}=200$ GeV
\cite{Arsene:2005mr,Bearden:2004yx,Bearden:2003hx} have not been
fitted within this model until now. The thermal parameters for the
three cases of the maximal RHIC collision energy have been taken
from the newer studies of the particle abundance ratios
\cite{Rafelski:2004dp,Barannikova:2005rw}.

The final results for the geometric parameters $\rho_{max}$ and
$\tau$ are gathered in Table~\ref{Table1} together with the
corresponding values of $\beta_{\perp}^{max}$ and $\chi^{2}$/NDF
for each centrality class characterized by the number of
participants $N_{part}$. All fits are statistically significant
beside the most peripheral bins of the PHENIX measurements.

\begin{table}
\caption{\label{Table1} Values of the geometric parameters of the
model fitted to the RHIC final data on the $p_{T}$ spectra of
identified charged hadrons
\protect\cite{Adcox:2003nr,Adler:2003cb,Adams:2003xp,Arsene:2005mr,Bearden:2004yx,Bearden:2003hx}.
Values of the thermal parameters are taken from the quoted
references. }
\begin{tabular}{cccccc} \hline Collision case &
$N_{part}$ & $\rho_{max}$ & $\tau$ & $\beta_{\perp}^{max}$ &
$\chi^{2}$/NDF \cr & & [fm] & [fm] & & \cr \hline PHENIX at &
347.7 & 6.50$\pm$0.27 & 8.23$\pm$0.23 & 0.62 & 0.52 \cr
$\sqrt{s_{NN}}=130$ GeV: & 271.3 & 5.99$\pm$0.21 & 7.29$\pm$0.18 &
0.63 & 0.46 \cr $T = 165$ MeV & 180.2 & 5.08$\pm$0.18 &
6.34$\pm$0.15 & 0.63 & 0.49 \cr $\mu_{B} = 41$ MeV & 78.5 &
3.59$\pm$0.15 & 4.81$\pm$0.13 & 0.60 & 0.74 \cr
\protect\cite{Florkowski:2001fp} & 14.3 & 1.68$\pm$0.19 &
3.14$\pm$0.22 & 0.47 & 1.32 \cr \hline PHENIX at & 351.4 &
8.46$\pm$0.10 & 8.84$\pm$0.08 & 0.69 & 0.80 \cr
$\sqrt{s_{NN}}=200$ GeV: & 299.0 & 7.99$\pm$0.10 & 8.23$\pm$0.08 &
0.70 & 0.61 \cr $T = 155.2$ MeV & 253.9 & 7.54$\pm$0.10 &
7.67$\pm$0.08 & 0.70 & 0.48 \cr $\mu_{B} = 26.4$ MeV & 215.3 &
7.11$\pm$0.10 & 7.17$\pm$0.07 & 0.70 & 0.48 \cr
\protect\cite{Rafelski:2004dp} & 166.6 & 6.45$\pm$0.09 &
6.47$\pm$0.07 & 0.71 & 0.58 \cr
 & 114.2 & 5.57$\pm$0.08 & 5.63$\pm$0.06 & 0.70 & 0.77
\cr
 & 74.4 & 4.68$\pm$0.07 & 4.85$\pm$0.06 & 0.69 & 1.05
\cr
 & 45.5 & 3.83$\pm$0.07 & 4.16$\pm$0.05 & 0.68 & 1.13
\cr
 & 25.7 & 2.99$\pm$0.06 & 3.47$\pm$0.05 & 0.65 & 1.41
\cr
 & 13.4 & 2.22$\pm$0.06 & 2.78$\pm$0.05 & 0.62 & 1.55
\cr
 & 6.3 & 1.71$\pm$0.06 & 2.40$\pm$0.05 & 0.58 & 1.40
\cr \hline STAR at & 352 & 9.22$\pm$0.18 & 7.13$\pm$0.06 & 0.79 &
0.29 \cr $\sqrt{s_{NN}}=200$ GeV: & 299 & 8.40$\pm$0.17 &
6.83$\pm$0.06 & 0.78 & 0.27 \cr $T = 160.0$ MeV & 234 &
7.57$\pm$0.15 & 6.33$\pm$0.06 & 0.77 & 0.23 \cr $\mu_{B} = 24.0$
MeV & 166 & 6.50$\pm$0.14 & 5.86$\pm$0.06 & 0.74 & 0.30 \cr
\protect\cite{Barannikova:2005rw} & 115 & 5.52$\pm$0.12 &
5.37$\pm$0.06 & 0.72 & 0.27 \cr
 & 76 & 4.66$\pm$0.11 & 4.91$\pm$0.06 & 0.69 & 0.27
\cr
 & 47 & 3.87$\pm$0.10 & 4.40$\pm$0.06 & 0.66 & 0.35
\cr
 & 27 & 3.07$\pm$0.09 & 3.94$\pm$0.06 & 0.61 & 0.46
\cr
 & 14 & 2.37$\pm$0.08 & 3.32$\pm$0.06 & 0.58 & 0.87
\cr \hline BRAHMS at & 357 & 8.75$\pm$0.16 & 8.38$\pm$0.11 & 0.72
& 0.50 \cr $\sqrt{s_{NN}}=200$ GeV: & 328 & 8.50$\pm$0.15 &
8.08$\pm$0.10 & 0.72 & 0.52 \cr $T = 155.2$ MeV & 239 &
7.52$\pm$0.13 & 7.28$\pm$0.09 & 0.72 & 0.46 \cr $\mu_{B} = 26.4$
MeV & 140 & 6.29$\pm$0.12 & 6.20$\pm$0.09 & 0.71 & 0.36 \cr
\protect\cite{Rafelski:2004dp} & 62 & 4.42$\pm$0.12 &
4.95$\pm$0.10 & 0.67 & 0.61 \cr \hline
\end{tabular}
\end{table}

\begin{figure}
\begin{center}{
{\epsfig{file=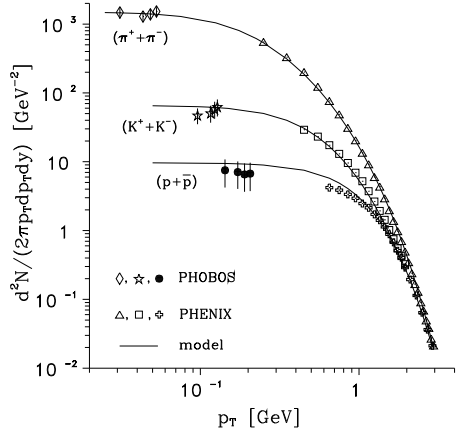,width=6cm}} }\end{center}
\caption{\label{Fig.1} Invariant yields as a function of $p_{T}$
for RHIC at $\sqrt{s_{NN}}=200$ GeV. The PHOBOS data are for the
$15 \%$ most central collisions with the error bars expressed as
the sum of the systematic and statistical uncertainties
\protect\cite{Back:2004zx}. The corresponding PHENIX data are
presented as the averages of the invariant yields for the $0-5
\%$, $5-10 \%$ and $10-15 \%$ centrality bins with no errors
given. Lines are the appropriate predictions of the
single-freeze-out model. }
\end{figure}
%%%%%%%%%%%%%%%%%%%%

To check the accuracy of the model predictions, the invariant
distributions given by Eq.~(\ref{Cooper2}) are calculated down to
the low-$p_{T}$ region (0.03-0.05 GeV for pions, 0.09-0.13 GeV for
kaons and 0.14-0.21 GeV for protons and antiprotons) of the PHOBOS
measurements at $\sqrt{s_{NN}}=200$ GeV \cite{Back:2004zx}. The
results are depicted in Fig.~\ref{Fig.1}. As one can see
predictions agree very well with the low-$p_{T}$ data.

\begin{figure}
\begin{center}{
{\epsfig{file=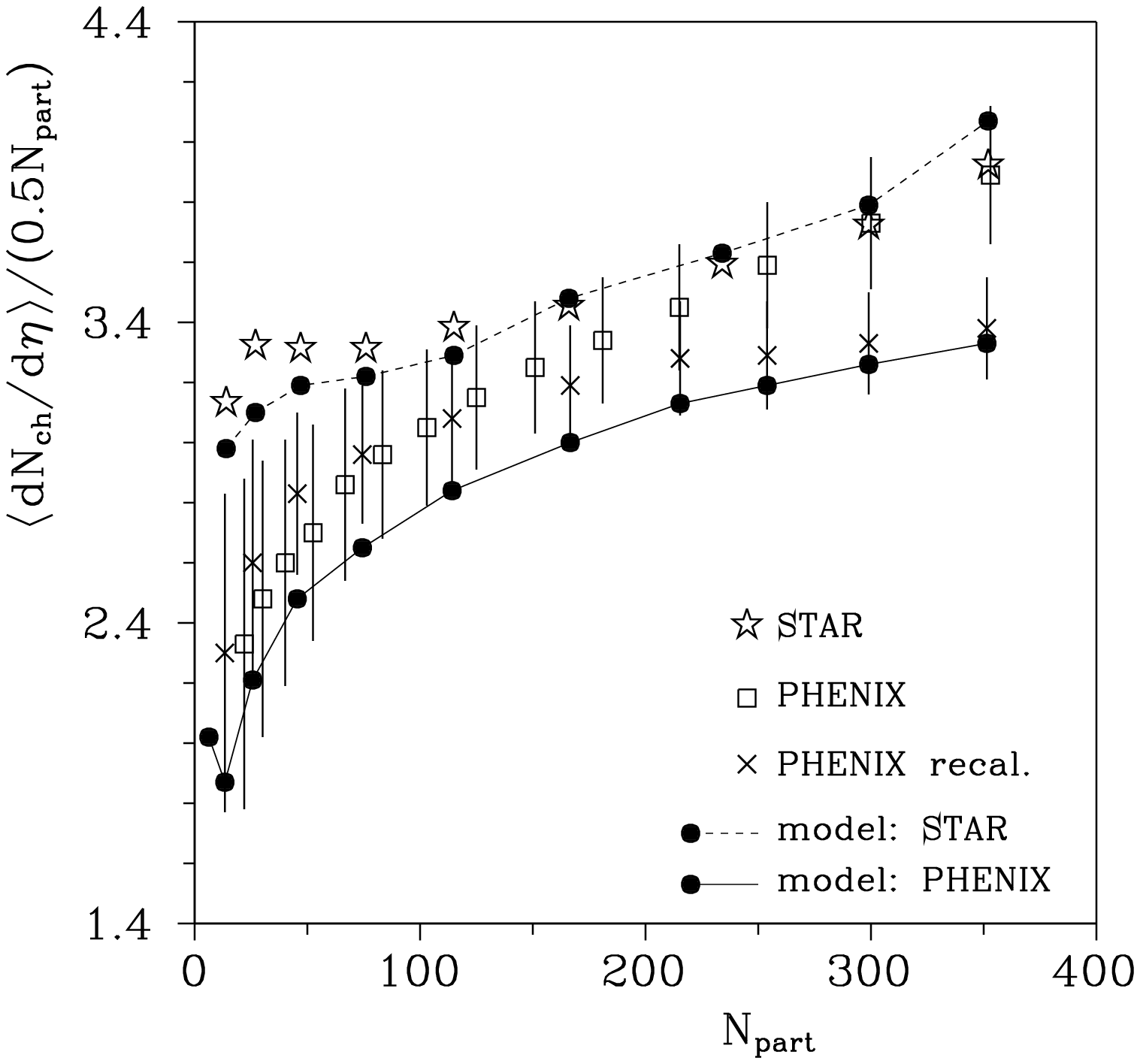,width=6cm}} }\end{center}
\caption{\label{Fig.2} $dN_{ch}/d\eta$ per pair of participants
versus $N_{part}$ for RHIC at $\sqrt{s_{NN}}=200$ GeV. The
original PHENIX data are from \protect\cite{Adler:2004zn}, whereas
the recalculated PHENIX data are from summing up the integrated
charged hadron yields delivered in \protect\cite{Adler:2003cb}.
Also the STAR data are depicted with no errors given as in the
source paper \protect\cite{Adams:2003xp}. The lines connect the
results and are to guide the eye. }
\end{figure}
%%%%%%%%%%%%%%%%

\begin{figure}
\begin{center}{
{\epsfig{file=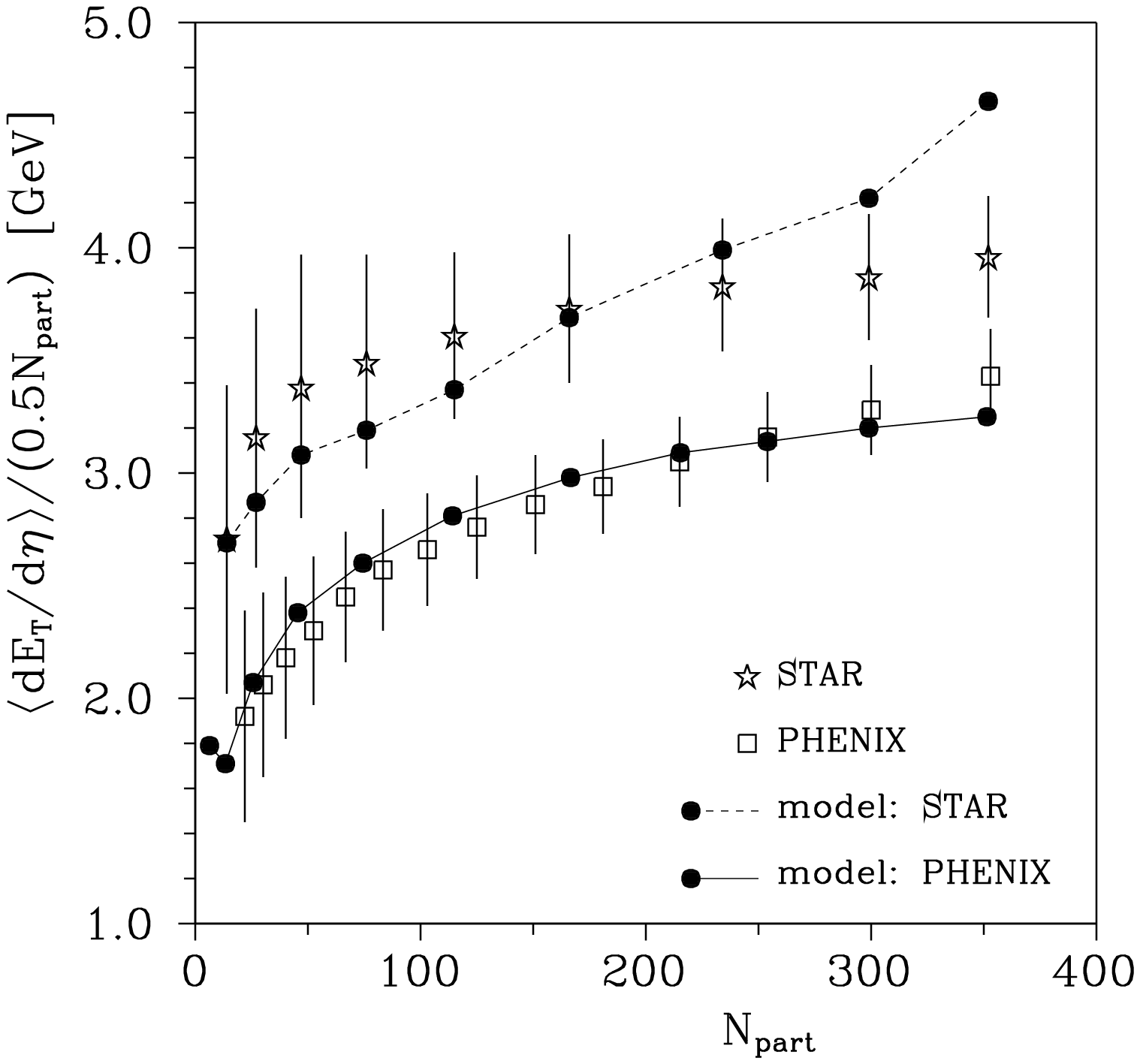,width=6cm}} }\end{center}
\caption{\label{Fig.3} $dE_{T}/d\eta$ per pair of participants
versus $N_{part}$ for RHIC at $\sqrt{s_{NN}}=200$ GeV. The PHENIX
data are from \protect\cite{Adler:2004zn} but the original STAR
data from \protect\cite{Adams:2004cb} have been rescaled to
$\eta=0$, see \protect\cite{Prorok:2005uv} for more explanations.
The lines connect the results and are to guide the eye. }
\end{figure}
%%%%%%%%%%%%%%

In Fig.~\ref{Fig.2} predictions for $dN_{ch}/d\eta\vert_{mid}$ per
participating pair as a function of centrality are presented for
$\sqrt{s_{NN}}=200$ GeV. In the STAR case results agree well with
the data. In the PHENIX case results agree within errors with the
data from the summing up of the integrated charged hadron yields
\cite{Adler:2003cb}, but they are significantly below the
straightforward PHENIX measurements \cite{Adler:2004zn}. This
reflects the observed discrepancy between the directly measured
$dN_{ch}/d\eta$ and $dN_{ch}/d\eta$ expressed as the sum of the
integrated charged hadron yields \cite{Chujo:2002bi}.

In Fig.~\ref{Fig.3} the estimates of $dE_{T}/d\eta\vert_{mid}$ per
participating pair are shown as a function of centrality for
$\sqrt{s_{NN}}=200$ GeV. The predictions agree well with the data,
only the most central point of the STAR case is substantially
overestimated ($\approx 17\%$).

\begin{figure}
\begin{center}{
{\epsfig{file=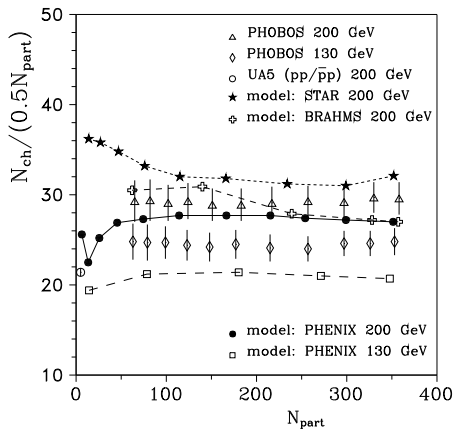,width=6cm}} }\end{center}
\caption{\label{Fig.4} $N_{ch}$ per pair of participants versus
$N_{part}$ for RHIC at $\sqrt{s_{NN}}=130$ and $200$ GeV. The
PHOBOS data are from \protect\cite{Back:2003xk} and the
$pp/\bar{p}p$ data point of the UA5 measurement is from Fig.39.5
in \protect\cite{Hagiwara:fs}. The lines connect the results and
are to guide the eye. }
\end{figure}

The predictions for the total charged-particle multiplicity per
participating pair are presented in Fig.~\ref{Fig.4}. One can see
that estimated values are roughly constant within the range of the
PHOBOS measurement \cite{Back:2003xk}, \ie $N_{part} \approx
60-360$. Only $\approx 10 \%$ deviation from this constancy can be
observed in the BRAHMS case. Also the predicted (absolut) values
agree with the data within $\approx 10 \%$.

%%%%%%%%%%%%%%

\section{Conclusions}

Global variables like the transverse energy density, the charged
particle multiplicity density and the total multiplicity of
charged particles have been estimated in the framework of the
single-freeze-out model for the Au-Au collisions at
$\sqrt{s_{NN}}=130$ and $200$ GeV. The estimates are based on the
fits of the parameters of the model which are done in the first
stage. While the thermal parameters have been taken from
independent fits to the particle yield ratios, the geometric
parameters have been determined from the final data on the
$p_{T}$-spectra of identified charged hadrons. The consistent
picture of the freeze-out has been obtained within the model since
the global variables are measured independently of identified
hadron spectroscopy and their predictions agree fairly well with
the data. It should be stressed that the model reproduces the
centrality independence of the total charged-particle multiplicity
per participating pair and the predicted values agree with the
measured ones within $\approx 10 \%$. This is surprising since
geometric parameters have been fitted to spectra measured at
midrapidity, but the total charged-particle multiplicity is
measured in the whole rapidity range.

The author would like to thank Jan Rafelski for very helpful
comments and discussions. This work was supported in part by the
Polish Committee for Scientific Research under Contract No. KBN 2
P03B 069 25.

\end{document}